\documentstyle[anta, twoside]{article}

\input{psfig.sty}

\def\Journal#1#2#3#4{(#1) {#2} {\bf #3}, #4}

\def\AAp{\em Astron. Astrophys.}
\def\AJ{\em Astron.~J.}
\def\ApJ{\em Astrophys.~J.}

\def\ApJSS{\em Astrophys.~J. Suppl.}

\def\AAS{\em Astron. Astrophys. Suppl.}

\def\MNRAS{\em Mon. Not. R.~Astron. Soc.}

\def\ApSS{\em Astrophys. Space. Sci.}
\begin{document}

\markboth{M. Johnston-Hollitt, C. P. Hollitt \& R. D.
Ekers}{Statistcial Analysis of Extra-galactic Rotation Measures}

\thispagestyle{plain}
\setcounter{page}{13}

\title{Statistical Analysis of Extra-galactic Rotation Measures \\}

%

%
 \author{M. Johnston-Hollitt$^1$, C. P. Hollitt$^2$ and R. D. Ekers$^3$}

\address{$^1$ Leiden Observatory,
             P.O. Box 9513 RA-2300 Leiden, The Netherlands \\
        $^2$ Department of Physics, School of Chemistry and Physics,\\
             University of Adelaide, Adelaide SA 5005, Australia\\
        $^3$ Australia Telescope National Facility,\\
             P.O. Box 76, Epping NSW 1710, Australia
 }


\maketitle

\abstract{We have performed a statistical analysis of a sample of 1100
extra-galactic rotation measures (RMs) obtained from the literature. Using
a subsample of approximately 800 reliable RMs we compute a rotation
measure sky and determine reliable large scale features for the line of
sight Galactic magnetic field. We find that the influence of the
Milky Way can be seen up to roughly $30\degr$ on either side of
the Galactic plane. Furthermore we observe an excess of RM on spatial
scales between $30\degr$ and $50\degr$ in the region of the Galactic
Plane. Additionally, the support for a bisymmetric spiral Galactic
magnetic field is significantly reduced in our analysis.}

\section{Introduction}

Magnetic fields are assumed to be pervasive throughout the Universe on
all scales, from the fields surrounding planets right up to fields in
the intracluster and intergalactic media. In recent years the role of
magnetic fields in both galactic and extra-galactic regimes has gained
increased attention across many astrophysical disciplines. For example,
the magnetic field is a key factor in studies of large-scale structure
formation, galaxy and star formation, and cosmic ray generation. In
particular, the Galactic magnetic field has been studied since the late
seventies with a variety of techniques. While it is clear that magnetic
field research has progressed considerably in this time, the mostly
indirect measurement techniques have meant that it has been difficult
to address many basic issues. Questions as to how strong the Galactic
magnetic field is, how uniform it is, what the seeding and
amplification mechanisms are, and, most importantly, what the
contribution is to the energy density of the galactic medium
remain topics of animated debate.

One of the ways in which the Galactic magnetic field can be examined
is through analysis of the rotation measures obtained for background
extra-galactic sources. This gives information on the line-of-sight
Galactic magnetic field and is complementary to results obtained from
other techniques such as pulsar dispersion measures and mapping of
the diffuse polarised emission of the Galaxy.

We present a statistical analysis of the RM sky as derived from a sample
of extra-galactic RMs given in the literature and use these data to
generate an interpolated map of the RM sky to give some insight
into properties of the large-scale Galactic magnetic field.

\section{RM Sample and Interpolated All-sky Mapping}

Over 1000 extra-galactic RMs taken from several catalogues
(Tabara \& Inoue, 1980; Simard-Normandin et al., 1981; Broten et al.,
1988; Hennessey et al., 1989; Rudnick \& Jones, 1983; Lawler \&
Dennison, 1982) were initially examined. Only those with a reliable RM
fit over at least three wavelengths were selected. In the case of
sources appearing in more than one reference, the more reliable fit was
used. This produced a final catalogue of 820 sources which was utilised
in a two-stage process.

\begin{figure}[htbp]
   \vspace{-0.65cm}
   \centerline{\psfig{figure=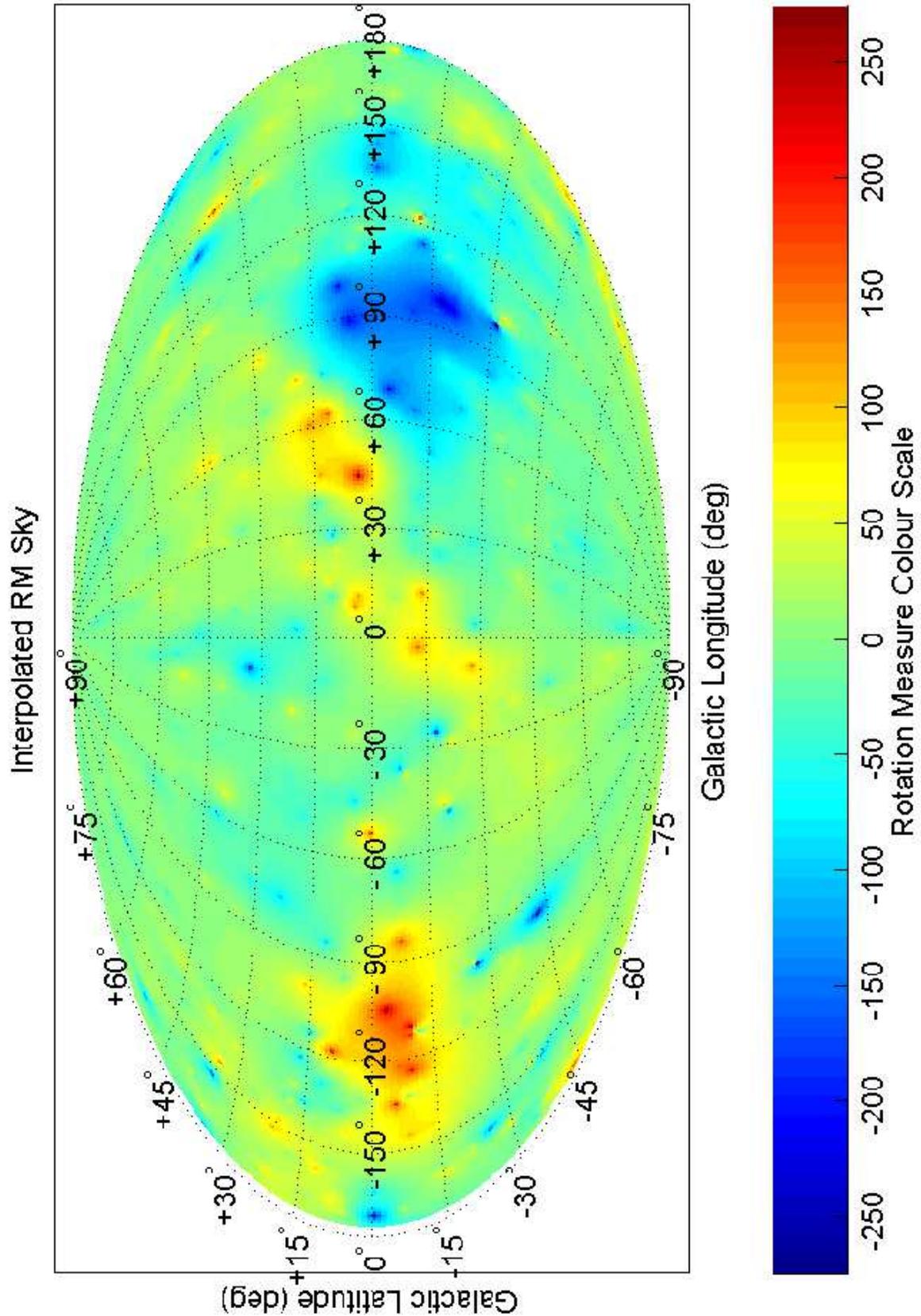,height=22.5cm,angle=0}}
   \caption[Estimated Rotation Measure Map]
           {Interpolated All-sky Rotation Measure Map, created from
            over 800 published RM values calculated for extra-galactic
            sources. The resolution is one pixel per square degree.}
 \label{fig:rm_interpcross}
\end{figure}

First, RMs projected through lines of sight through galaxy clusters were
removed as it has been shown that they would be contaminated by passage
through the cluster magnetic field (Clarke, 2000; Johnston-Hollitt, 2003).
Next a culling algorithm removed sources for which there was a
three-sigma deviation from the local median modulus RM. This was
similar to previous small-scale Galactic RM estimation techniques
(Hennessey et al., 1989; Athreya et al., 1997) but rather than using a
defined radius about an individual point the algorithm tests the
population of at least the nine nearest neighbours. Moreover, unlike
previous techniques, we estimate sigma from the median modulus value
rather than simply the mean. The effect of the culling was to remove
19 sources of extremely high intrinsic RM. The estimation of
line-of-sight RMs across the entire sky was then performed by obtaining a
convergent solution to the 2-dimensional Poisson's equation.
Unfortunately the source density of this dataset is approximately
0.013 sources per square degree and so the resolution of the resultant map
was set to be no finer than one pixel per square degree.
Figure \ref{fig:rm_interpcross} shows the resultant estimated all-sky RM map
in Galactic coordinates. This clearly shows the strong correlation between
RM and distance from the Galactic plane.

\section{Structure Analysis}

With such a dataset and all-sky RM map it is possible to statistically
determine properties of the RM sky, such as the region in which our
Galaxy significantly affects RM measurements. We investigated the data
in several ways including a Fourier analysis of the RM sky at various
Galactic latitudes and the source distribution for extra-galactic RMs
beyond the region of Galactic influence.

\subsection{Fourier Analysis}

A Fourier analysis of the power at different spatial scales
in the interpolated RM map was performed within different strips along the
Galactic plane with increasing latitude. The power spectra within
increasing intervals from the Galactic plane were examined starting
from $\pm15\degr$ and increasing to $\pm60\degr$ in steps of $15\degr$.
An excess of RM power was seen at $30\degr$, $46\degr$ and $50\degr$
in the small interval of $\pm15\degr$ about the plane. In this region
closest to the Galactic plane there was also marginal evidence for an
excess at around $80\degr$. In the interval $\pm30\degr$ about the
plane an excess was seen on spatial scales of $30\degr$ and $50\degr$.
In the intervals above $30\degr$ there was only very marginal
evidence for some excess at around $50\degr$ in the $\pm45\degr$
interval.

These results show that the Galactic magnetic field significantly
influences RMs of extra-galactic sources that lie within an area
roughly $30\degr$ either side of the Galactic plane. A comparative
study of excess RM above and below the Galactic plane was also
conducted. From this second analysis we obtain marginal evidence that
the area of influence of the Galaxy extends further in the south.
However, as there are much fewer actual RM measurements in this region
of the sky this could well be the affect of poor and uneven sampling
(as compared to the rest of the sky).

Reassuringly, results from this method are consistent with the
large-scale positive and negative RM regions seen in the wavelet
analysis study of Frick et al.\ (2001), which makes use of a similar
dataset but does not make any attempt to remove RM values that are
dubious or where a source has more than one value. In comparison to
Frick et al.\ (2001) who obtain structures on spatial scales of roughly
30\degr --$45\degr$ and $76\degr$, we find excess RM power on
scales between $30\degr$ and $50\degr$ within a region of $30\degr$
either side of the Galactic plane. Moreover, we find weak evidence of
an excess at a scale of around $80\degr$.

\subsection{Global Field Examination}

Comparison of the large-scale RM features with the position of the
spiral arms as deduced from electron density models
(Taylor \& Cordes, 1993) demonstrates that these features are likely to
be correlated with the magnetic field in the interarm region between
the Perseus and Sagittarius spiral arms. The number and position
of field reversals in the Milky Way are critical to distinguish the
global field as either axisymmetric or bisymmetric (Vall\'{e}e, 1996)
and proponents of various models often have widely varying
interpretations of the global direction of each arm and its associated
interarm regions (Vall\'{e}e, 1991; Clegg et al., 1992; Han et al.,
1994; Rand et al., 1994; Vall\'{e}e, 1996). It is generally agreed that
the field rotates in a clockwise direction in the interarm region
between the Perseus and Sagittarius--Carina arms and in an
anti-clockwise direction in the region between the Sagittarius--Carina
and Scutum--Crux arms. In addition, the weight of the literature
supports a clockwise rotation in the interarm region between the
Scutum-Crux and Norma arms. However, the field direction between the
other spiral arms in the outer and inner parts of the galaxy is an area
of contention.

The interpolated data appear consistent with a clockwise field
direction beyond the Perseus spiral arm, in the region between the
Perseus and Perseus +I arm, however, this is inconsistent with pulsar
dispersion measures (Han et al., 1994) but agrees with models including
other RM data (Vall\'{e}e, 1996).

The interpolated data are not conclusive for other regions. In the
regions corresponding to both the Perseus/Sagittarius--Carina interarm
region the interpolated data show both positive and negative features.
In the area between the Sagittarius--Carina and Scutum--Crux arms these
data show only a positive RM region. One notable feature in the
interpolated map from the new dataset is the lack of the alternating
positive-negative-positive-negative RM average in the four quadrants of
the sky which had been claimed in previous work (Han et al., 1997).
As the alternating positive-negative signature is believed to give
evidence for a bisymmetric field structure in our Galaxy, the lack of
the expected positive average RM in the region between $0\degr\leq l \leq
180\degr$ and $0\degr\leq b \leq 90\degr$ puts the bisymmetric model in
some doubt. We note that Frick et al.\ (2001) also find only marginal
evidence for this signature. This suggests this feature is highly
sensitive to even small changes in the dataset used and more RMs will
be required to settle this point.

\subsection{Source Statistics}
\label{sourcestats}

\begin{figure}[htbp]%
   \centerline{\psfig{figure=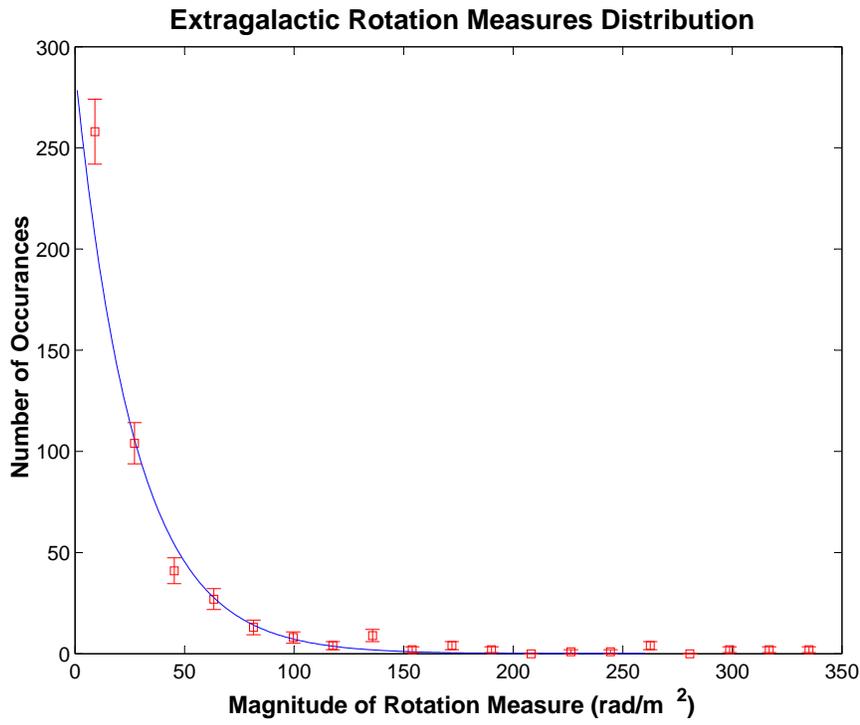,height=9.5cm}}
   \caption {Plot of the extra-galactic RM distribution for sources
greater than $30\degr$ from the Galactic plane. The boxes represent
the data, while the line is the chi-squared fit to the data.}
 \label{fig:rm_stats}
\end{figure}

\begin{figure}[htbp]
   \centerline{\psfig{figure=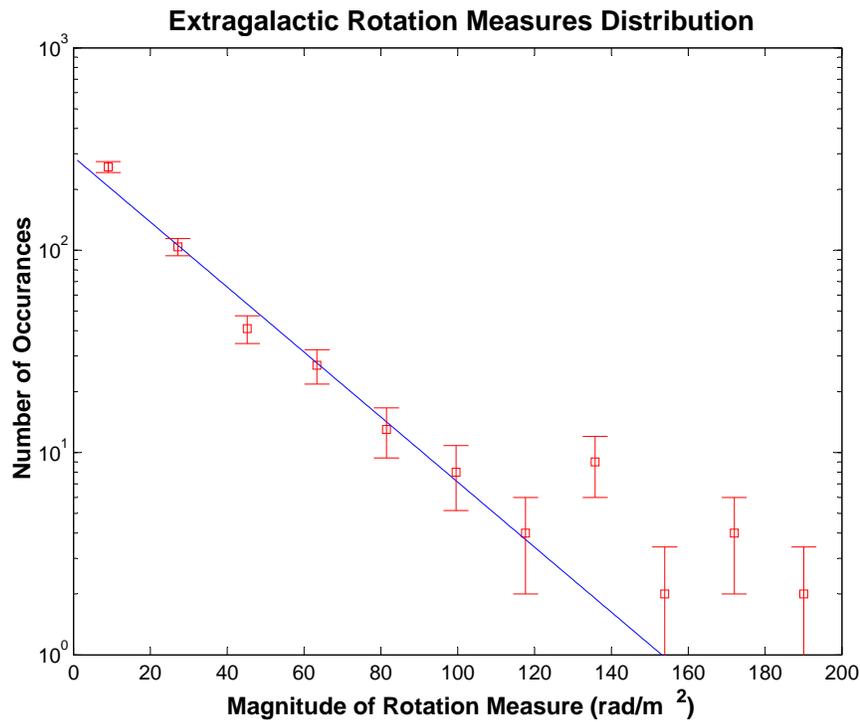,height=9.5cm}}
   \caption{Plot of the extra-galactic RM distribution for sources
greater than $30\degr$ from the Galactic plane shown as a log--linear
plot. The boxes represent the data, while the line is the chi-squared
fit to the data.}
 \label{fig:rm_statslog}
\end{figure}

In order to investigate the statistical behaviour of the RM distribution,
subsets of the data at various distances from the Galactic plane were
examined prior to removing the high intrinsic RM sources. In particular,
the region greater than $30\degr$ from the Galactic plane was heavily
investigated as this had previously been shown to be beyond the
influence of the Galactic field (see Section 3.1). The standard
deviation of 474 extra-galactic RMs at greater than $30\degr$ from
the Galactic plane was found to be 10~rad~m$^{-2}$ which is consistent
with more localised calculations taken at such high Galactic latitudes
(Athreya et al., 1997; Clarke, 2000). Furthermore, it was discovered
that at high galactic latitudes the source distribution follows an
exponential. This result is both interesting and unexpected as at these
latitudes one expects little or no contribution to the RM from the
magnetic field of the Galaxy, suggesting that the exponential
distribution must either be a product of internal rotation in the
extra-galactic sources or propagation through different magnetized
cells in the interstellar medium. It was previously thought that this
distribution would be Gaussian. This is an interesting and important
result which implies that the occurrence of intrinsically high RMs is
currently being underestimated. Figure \ref{fig:rm_stats}
shows the distribution of RM at $\vert b\vert \geq 30\degr$ for the
Galactic plane overlaid with an exponential fit to the data. Figure
\ref{fig:rm_statslog} shows the same data but with a log--linear plot.
Chi-squared testing shows this data to be exponential of the form
Number of occurrences = A exp($-$0.037 $\times$ RM) (where A is a
scaling constant in this case A=289) to greater than the 99.9\%
confidence level. In comparison, the distribution obtained from all
data, i.e. including those RMs seen on lines of sight through the
Galaxy shows a marked deviation from the exponential fit, especially
for sources with modulus RMs greater than 50~rad~m$^{-2}$.

\section{Conclusions}

We have presented an analysis of the currently available reliable
population of extra-galactic rotation measures. From this we find that
the influence of the Galactic magnetic field is statistically
significant out to $\pm 30\degr$ from the Galactic plane and that it
has excess power on spatial scales of $30\degr$, $46\degr$ and $50\degr$.
This agrees well with alternative analyses. We further find that the
current data do not suggest the alternating positive-negative
average RM values in each Galactic quadrant required to support a
bisymmetric magnetic field for the Milky Way. Furthermore, we show that
the population of high Galactic RMs presumed to be unaffected by the
magnetic field of our Galaxy follows an exponential distribution to
above 99.9\% confidence.

With the completion, or near completion, of major polarimetric surveys
such as the Southern Galactic Plane Survey and the Canadian Galactic
Plane Survey, a wealth of new information on the magnetic field
structure of our galaxy will soon become available. The expected boost
in available data for this work is a factor of 5--10. Tantalising
preliminary results have recently appeared, giving new insight into
both the RM structure of the sky as seen through the Galaxy and the
role of the magnetic field in diffuse Galactic plasma (Gaensler et
al., 2001; Brown et al., 2001). As these new data become available,
better modelling and possibly even subtraction of the effect of the
local field will be possible. With new instruments such as the SKA it
may even be possible to completely disentangle the propagation effects
on RM data and see, for the first time, the 3D magnetic structure of
the Universe. Thus, the interpolated technique presented here
should continue to be useful in evaluating the rotation measure sky.

\section*{Acknowledgments}
We would like to thank Dr. JinLin Han and Dominic Schnitzeler for useful
discussions on the early stages of the work. In addition, M. J.-H. and
C. P. H. would like to thank the conference organisers for selecting such
a beautiful and interesting environment.

\section*{References}\noindent

\references

Athreya R.M., Kapahi V.K., McCarthy P.J., van Breugel W.
   \Journal{1997}{\AAp}{329}{809}.

Broten N.W., MacLeod J.M., Vall\'{e}e J.P. \Journal{1988}{\ApSS}{141}{303}.

Brown J.C, Taylor A.R. \Journal{2001}{\ApJ}{563}{L31}.

Clarke T.E. (2000) {\em Probing Magnetic Fields in Clusters of
  Galaxies} (Ph.D. Thesis, University of Toronto).

Clegg A.W., Cordes J.M., Simonetti J.H., Kulkami S.R.
  \Journal{1992}{\ApJ}{386}{143}.

Frick P., Stepanov R., Shukurov A., Sokoloff D.
  \Journal{2001}{\MNRAS}{325}{649}.

Gaensler B.M., Dickey J.M., McClure-Griffiths N.M., Green A.J., Wieringa M.H.,
  Haynes R.F. \Journal{2001}{\ApJ}{549}{959}.

Han J.-L., Qiao G.J. \Journal{1994}{\AAp}{288}{759}.

Han J.-L., Manchester R.N., Berkhuijsen E.M., Beck R.
  \Journal{1997}{\AAp}{322}{98}.

Hennessey G.S., Owen F., Eilek J. \Journal{1989}{\ApJ}{347}{144}.

Johnston-Hollitt M. (2003) {\em Detection of Magnetic Fields and
  Diffuse Radio Emission in Abell 3667 and other Rich Southern Clusters
  of Galaxies} (Ph.D. Thesis, University of Adelaide).

Lawler J.M., Dennison B. \Journal{1982}{\ApJ}{252}{81}.

Rand R.J., Lyne A.G. \Journal{1994}{\MNRAS}{268}{497}.

Rudnick L., Jones T. \Journal{1983}{\AJ}{88}{518}.

Simard-Normandin M., Kronberg P.P., Button S. \Journal{1981}{\ApJSS}{45}{97}.

Tabara H., Inoue, M. \Journal{1980}{\AAS}{39}{379}.

Taylor J.H., Cordes J.M. \Journal{1993}{\ApJ}{411}{674}.

Vall\'{e}e J.P. \Journal{1991}{\ApJ}{366}{450}.

Vall\'{e}e J.P. \Journal{1996}{\AAp}{308}{433}.

\end{document}